\documentclass[traditabstract,longauth]{aa}

\usepackage{hyperref}
\hypersetup{
    bookmarks=true,                     
    unicode=false,                      
    pdftoolbar=true,                    
    pdfmenubar=true,                    
    pdffitwindow=true,                  
    pdftitle={Spiral-like patterns},    
    pdfauthor={Author},                 
    pdfsubject={Spiral-like patterns},  
    pdfnewwindow=true,                  
    pdfkeywords={keywords},             
    colorlinks=true,                    
    linkcolor=blue,                     
    citecolor=blue,                     
    filecolor=blue,                     
    urlcolor=blue                       
}

\usepackage{natbib}
\usepackage{lscape}
\usepackage[section]{placeins}
\usepackage{wasysym}
\usepackage{graphics}
\usepackage{graphicx}
\usepackage{txfonts}
\usepackage{color}
\usepackage{pgfkeys}
\usepackage[]{tcolorbox}
\usepackage{times}
\usepackage{mathtools}
\usepackage[T1]{fontenc}
\newcommand{\changefont}[3]{\fontfamily{#1}\fontseries{#2}\fontshape{#3}\selectfont}
\newfont{\vcap}{cmssdc10 scaled 1000}
\newfont{\lcap}{cmssdc10 scaled 1100}
\newfont{\mcap}{cmssdc10 scaled 1400}
\newfont{\nlx}{cmssdc10 scaled 900}
\newfont{\xnlx}{cmssdc10 scaled 800}
\newfont{\hvss}{cmssdc10 scaled 1540}

\newcommand{\sbb}{mag/$\sq\arcsec$}

\def\rr{{\sl R}$^{\star}$}
\def\ha{H$\alpha$}
\def\ewha{EW(\ha)}

\def\lo3hb{$\log$([O\,{\sc iii}]/H$\beta$)}
\def\ln2ha{$\log$([N\,{\sc ii}]/H$\alpha$)}
\def\tauha{$\tau$}
\def\tauha_ext{$\tau$}

\def\pAGBmax{{\sc ew}$_{\star}^+$}

\def\rr{{\sl R}$^{\star}$}

\def\eqan{\begin{equation}}
\def\eqen{\end{equation}}

\def\h1{\ion{H}{i}}
\def\h2{\ion{H}{ii}}
\def\reff{{\it R$_{\rm eff}$}}

\def\rr{{\sl R}$^{\star}$}

\def\ha{H$\alpha$}
\def\ewha{EW(\ha)}

\def\hb{H$\beta$}

\def\o5007{[O {\sc iii}] $\lambda$5007}

\def\n2ha{[N\,{\sc ii}]/H$\alpha$}
\def\ln2ha{$\log$([N\,{\sc ii}]/H$\alpha$)}
\def\tn2ha{[N\,{\sc ii}]${\scriptstyle 6584}$/H$\alpha$}
\def\tln2ha{$\log$([N\,{\sc ii}]${\scriptstyle 6584}$/H$\alpha$)}
\def\o3hb{[O\,{\sc iii}\,]/H$\beta$}
\def\lo3hb{$\log$([O\,{\sc iii}]/H$\beta$)}
\def\to3hb{[O\,{\sc iii}]${\scriptstyle 5007}$/H$\beta$}
\def\tlo3hb{$\log$([O\,{\sc iii}]${\scriptstyle 5007}$/H$\beta$)}
\def\e16{$10^{-16}~{\rm erg\,s^{-1}\,cm^{-2}}$}
\def\e17{$10^{-17}~{\rm erg\,s^{-1}\,cm^{-2}}$}
\def\tauha{$\tau$}
\def\tauha_ext{$\tau$}

\def\pAGBmax{$\mathrm{EW_{\star}^+}$}

\def\wim{\changefont{cmtt}{m}{it}wim\rm}

\def\ha{H$\alpha$}
\def\ewha{EW(\ha)}

\def\tauha{$\tau$}
\def\tauha_ext{$\tau$}

\def\pAGBmax{{\sc ew}$_{\star}^+$}

\def\isan{\nlx isan\rm} 
\def\mstar{${\cal M}_{\star}$}

\def\porto3d{\sc Porto3D\rm}
\def\p3d{\sc Porto3D\rm}
\def\P3D{\sc Porto3D\rm}
\def\SL{{\sc Starlight}\rm}
\newcommand\btab[5]{\begin{table}[#1]\label{#3}{\parbox{#4}{\caption{#2}}\rule[-0.5ex]{0cm}{0.5ex} }
\begin{tabular*}{#4}{#5} \label{#3} }

\newcommand{\etab}[3]{
\end{tabular*}

\vspace*{#1}
\begin{flushleft}
\parbox{#2}{\footnotesize #3}
\end{flushleft}
\end{table} }
\newcommand{\setab}[3]{
\end{tabular*}
\end{footnotesize}

\vspace*{#1}
\begin{flushleft}
\parbox{#2}{\footnotesize #3}
\end{flushleft}
\end{table} }

\def\eqan{\begin{equation}}
\def\eqen{\end{equation}}

\def\h1{\ion{H}{i}}
\def\h2{\ion{H}{ii}}
\def\reff{{\it R$_{\rm eff}$}}

\newcommand{\kmsec}{km~s$^{-1}$}
\newcommand{\msun}{M$_\odot$}

\def\rr{{\sl R}$^{\star}$}
\def\P25{{\sl R}$_{\rm SF}$}
\def\E25{{\sl R}$_{\rm host}$}

\def\mstar{${\cal M}_{\star}$}

\def\m5{${\cal M}_{\star,{\rm 5\,Gyr}}$}

\def\D4000{$D_{4000}$}

\def\ha{H$\alpha$}
\def\ewha{EW(\ha)}

\def\hb{H$\beta$}

\def\e16{$10^{-16}~{\rm erg\,s^{-1}\,cm^{-2}}$}
\def\e17{$10^{-17}~{\rm erg\,s^{-1}\,cm^{-2}}$}
\newcommand{\PutLabel}[3]{\put(#1,#2){#3}}
\newcommand {\aga} {\ {\raise-.5ex\hbox{$\buildrel>\over\sim$}}\ }
\newcommand {\ala} {\ {\raise-.5ex\hbox{$\buildrel<\over\sim$}}\ } 
\definecolor{orange}{rgb}{1,0.5,0}
\definecolor{mred}{rgb}{0.55,0.02,0.018}

\begin{document}

\titlerunning{Spiral-like star-forming patterns in CALIFA early-type galaxies}
\authorrunning{J.~M.~Gomes et~al.}

\title{Spiral-like star-forming patterns in CALIFA early-type galaxies
\thanks{Based on observations collected at the Centro Astron\'omico
Hispano Alem\'an (CAHA) at Calar Alto, operated jointly by the Max-Planck-Institut 
f\"ur Astronomie (MPIA) and the Instituto de Astrof\'isica de
Andaluc\'ia (CSIC)}}

\author{
J.~M.~Gomes\inst{\ref{inst1}}
\and P.~Papaderos\inst{\ref{inst1}}
\and J.~M.~V\'{\i}lchez\inst{\ref{inst2}}
\and C.~Kehrig\inst{\ref{inst2}}
\and J.~Iglesias-P\'aramo\inst{\ref{inst2},\ref{inst3}}
\and I.~Breda\inst{\ref{inst1}}
\and M.~D.~Lehnert\inst{\ref{inst4}}
\and S.~F.~S\'anchez\inst{\ref{inst5}}
\and B.~Ziegler\inst{\ref{inst6}}
\and S.~N.~dos Reis\inst{\ref{inst1}}
\and J.~Bland-Hawthorn\inst{\ref{inst7}}
\and L.~Galbany\inst{\ref{inst8},\ref{inst9}}
\and D.~J.~Bomans\inst{\ref{inst10},\ref{inst11}}
\and F.~F.~Rosales-Ortega\inst{\ref{inst12}}
\and C.~J.~Walcher\inst{\ref{inst13}}
\and R.~Garc{\'\i}a-Benito\inst{\ref{inst2}}
\and I.~M\'arquez\inst{\ref{inst2}}
\and A.~del Olmo\inst{\ref{inst2}}
\and M.~Moll\'a\inst{\ref{inst14}}
\and R.~A.~Marino\inst{\ref{inst15},\ref{inst16}}
\and C.~Catal\'an-Torrecilla\inst{\ref{inst17}}
\and R.~M.~Gonz\'alez Delgado\inst{\ref{inst2}}
\and \'A.~R.~L\'opez-S\'anchez\inst{\ref{inst18},\ref{inst19}}
\and the CALIFA collaboration
}
\offprints{jean@astro.up.pt}

\institute{
Instituto de Astrofísica e Ciências do Espa\c{c}o, Universidade do Porto,
Centro de Astrof{\'\i}sica da Universidade do Porto, Rua das Estrelas, 4150-762 Porto, 
Portugal\label{inst1}
\and
Instituto de Astrof\'isica de Andaluc\'ia (CSIC), Glorieta de la Astronom\'{\i}a s/n Aptdo. 3004, E18080-Granada, Spain\label{inst2}
\and
Estaci\'on Experimental de Zonas Aridas (CSIC), Ctra. de Sacramento s.n., La Ca\~nada, Almería, Spain\label{inst3}
\and
Institut d'Astrophysique de Paris, UMR 7095, CNRS, Universit\'{e} Pierre et Marie Curie, 98 bis boulevard Arago, 75014 Paris, France\label{inst4}
\and
Instituto de Astronom\'ia,Universidad Nacional Auton\'oma de Mexico, A.P. 70-264, 04510, M\'exico, D.F.\label{inst5}
\and
University of Vienna, T\"{u}rkenschanzstrasse 17, 1180 Vienna, Austria\label{inst6}
\and
Sydney Institute for Astronomy, University of Sydney, NSW 2006,
Australia\label{inst7}
\and
Millennium Institute of Astrophysics, Chile\label{inst8}
\and
Departamento de Astronom\'ia, Universidad de Chile, Casilla 36-D, Santiago, Chile\label{inst9}
\and
Astronomical Institute of the Ruhr-University Bochum\label{inst10}
\and
RUB Research Department Plasmas with Complex Interactions\label{inst11}
\and
Instituto Nacional de Astrof\'isica, \'Optica y Electr\'onica, Luis E. Erro 1, 72840 Tonantzintla, Puebla, Mexico\label{inst12}
\and
Leibniz-Institut f\"ur Astrophysik Potsdam (AIP), An der Sternwarte 16, D-14482 Potsdam, Germany\label{inst13}
\and
CIEMAT, Avda. Complutense 40, 28040 Madrid, Spain\label{inst14}
\and
CEI Campus Moncloa, UCM-UPM, Departamento de Astrof\'isica y CC. de la
Atm\'osfera, Facultad de CC. F\'isicas, Universidad Complutense de Madrid,
Avda. Complutense s/n, 28040 Madrid, Spain\label{inst15}
\and
Department of Physics, Institute for Astronomy, ETH Z\"urich, CH-8093
Z\"urich, Switzerland\label{inst16}
\and
Departamento de Astrof\'{\i}sica y CC. de la Atm\'{o}sfera, Universidad
Complutense de Madrid, E-28040, Madrid, Spain\label{inst17}
\and
Australian Astronomical Observatory, PO Box 915, North Ryde,
NSW 1670, Australia\label{inst18}
\and
Department of Physics and Astronomy, Macquarie University, NSW
2109, Australia\label{inst19}
}
\date{Received ?? / Accepted ??}
\abstract{
  Based on a combined analysis of SDSS imaging and CALIFA integral field
  spectroscopy data, we report on the detection of faint (24 $<$ $\mu_r$ \sbb\
  $<$ 26) star-forming spiral-arm-like features in the periphery of three
  nearby early-type galaxies (ETGs).  These features are of considerable
  interest because they document the still ongoing inside-out growth of some
  local ETGs and may add valuable observational insight into the origin and
  evolution of spiral structure in triaxial stellar systems.  A characteristic
  property of the nebular component in the studied ETGs, classified i+, is a
  two-radial-zone structure, with the inner zone that displays faint
  (\ewha$\simeq$1\AA) low-ionization nuclear emission-line region (LINER)
  properties, and the outer one (3\AA $<$\ewha$\la$20\AA) H{\sc ii}-region
  characteristics.  This spatial segregation of nebular emission in two
  physically distinct concentric zones calls for an examination of aperture
  effects in studies of type~i+ ETGs with single-fiber spectroscopic data.
} 

\maketitle
\markboth{Gomes et~al.}{Spiral-like star-forming patterns in CALIFA early-type galaxies}

\section{Introduction \label{intro}}
Studies of low-level star formation (SF) activity in early-type galaxies
(ETGs) are fundamental for advancing our understanding of the build-up history
and spectrophotometric evolution of these systems and elucidating the nature
of the dominant excitation source of their warm interstellar medium
(\wim). Particularly valuable in this respect are observational insights into
the spatial characteristics and temporal evolution of SF at the morphological
interface between lenticulars (S0) and ellipticals (E), since they can offer
key constraints on the origin of quenching (or, conversely, rejuvenation) in
ETGs, and its implications on the spectrophotometric evolution of their
structural components.

The origin of faint nebular emission in ETGs -- long revealed through
single-aperture and longslit spectroscopy
\citep[e.g.,][]{Phillips1986,dem84,kim89,tri91,ani10,YanBlanton2012},
narrow-band imaging \citep[e.g.,][]{Finkelmann2010,Kulkarni2014}, and more
comprehensively, with integral field spectroscopy \citep[IFS,
e.g.,][]{Sarzi06,Sarzi10,K12,P13,Singh13} -- continues to provide fodder for
an intense ongoing debate.  Low-level SF activity \citep[see, e.g.,][and
references therein]{Trager2000,Schawinski2007,Shapiro2010} offers an appealing
hypothesis for the excitation of the extended (extra-nuclear) \wim\ in ETGs,
as an alternative or supplementary mechanism to photoionization by an active
galactic nucleus \citep[AGN, e.g.,][]{Ho2008} or the evolved ($\geq 10^8$ yr)
post-AGB (pAGB) population \citep[e.g.,][]{tri91,bin94,sta08}, or excitation
by fast shocks \citep[e.g.,][]{DopitaSutherland1995}. Observational evidence
for SF in ETGs comes from combined near-UV, SDSS~$r$ and mid-IR color studies
\citep[e.g.,][]{Kaviraj07,GdP07,Ko14}, with 10--15\% of the stellar mass
(\mstar) in these systems assembling in a declining SF process since $z \simeq
1$ \citep{Kaviraj08}. The presence of a small fraction ($\sim$5.7\%) of blue
ETGs in the local universe with estimated star-formation rates (SFRs) between
0.5 and 50 \msun/yr \citep{Schawinski2009} lends additional support to this
notion.  That an appreciable fraction of nearby ETGs continues its build-up at
a very slow pace, with specific SFRs of $\la 10^{-11}$ yr$^{-1}$
\citep[e.g.,][]{Salim12}, is further indicated by multiwavelength studies that
combine single-fiber spectroscopy with GALEX/SDSS photometry
\citep[e.g.,][]{HuangGu09,Petty13,Pan14}.

However, our knowledge on the spatial distribution of SF in ETGs is still
fragmentary, even though a few detailed studies of individual systems or small
galaxy samples seem to unequivocally point to an increasing luminosity
contribution from young-to-intermediate-age stellar populations towards their
periphery.  This is suggested by, e.g., UV-optical color profiles of bluish
ETGs at the boundary of the green valley, which displays red centers and blue
outskirts \citep{Fang2012}.  UV-bright shells in the extranuclear component of
some nearby ETGs
\citep[e.g.,][]{Rampazzo07,Marino2011a,Marino2011b,Bresolin2013} underscore
this conclusion.  The prolonged build-up of the peripheral zones of ETGs is
also reflected in a radial decrease in the luminosity-weighted stellar age,
with $\sim$50\% of the $V$-band luminosity contributed by stars younger than 5
Gyr \citep[][hereafter G15b]{G15b}. Circumstantial evidence for low-level SF
in the outskirts of ETGs also comes from recent IFS studies \citep[][hereafter
P13]{P13}, which reveals that the \ha\ equivalent width (\ewha) of more than
one half of ETGs increases outwardly.

In the light of such results, a representative census of SF patterns in ETGs
appears to be of considerable interest because it may hold clues to the
structural evolution of these systems and the role of intrinsic and
environmental properties on the transformation of S0 into E galaxies.  For
example, a full understanding of the Butcher-Oemler \citep{BO78} effect in
galaxy clusters requires understanding of how the gradual cessation of SF
activity and photometric dimming of the progenitor disk component is
synchronized with the growth of the bulge and how this leads to a high
bulge-to-total ratio (B/T) present-day ETG. \citet{Johnston15} envisage a
two-stage formation process for the bulge of S0 galaxies with the first phase
completed early on and the second one fed by inflowing gas from the disk. This
takes place in tandem with the starvation of SF activity in the latter as the
result of ram pressure stripping. On the other hand, the evolutionary pathways
of ETGs in lower-density environments are less clear, leaving space for
various channels for the morphological S0$\rightarrow$E transition.

Regardless of the environmental properties of present-day ETGs and of whether
their SF activity is viewed from the perspective of quenching or rejuvenation
that is possibly driven by minor merges \citep[e.g.,][]{Mapelli15}, it is
important to explore how the amplitude and spatial properties of low-level SF
in these systems are related to their gas mass fraction and B/T ratio, how the
SFR surface density and stellar initial mass function (IMF) change as the
cold-gas supply is gradually depleted, and what intrinsic properties determine
whether SF is spatially organized in contiguous rims or spiral-arm-like
patterns, rather than in a diffuse and/or patchy collection of faint H{\sc ii}
regions.

IFS studies can obviously add key insights into this subject, permitting
simultaneous analysis of SF patterns and kinematics at the interface between
S0 and E galaxies.  A superb observational foundation for this is now provided
by the Calar Alto Legacy Integral Field Area survey
\citep[CALIFA,][]{Sanchez2012} -- the first IFS survey to map the entire
optical spectral range and spatial extent of a representative sample of the
Hubble sequence in the local universe.

Here we report on the detection of faint star-forming spiral-arm-like features
in three ETGs. The galaxies we study are similar to type~i ETGs -- systems
characterized by a nearly constant \ewha\ $\leq$ 3\AA\ throughout their
extranuclear component (P13) -- with the additional feature of a steep \ewha\
increase in their outermost periphery. This led us in \citet{G15a} to assign
them the notation i+ and motivated a more detailed analysis here.  In
Sect.~\ref{data} we briefly comment on the general properties of the i+ ETGs
and in Sect.~\ref{results} demonstrate, through a combined analysis of SDSS
imaging and CALIFA IFS data, that the \ewha\ excess in the periphery of these
systems comes from low-level SF that is associated with an ongoing inside-out
growth process.  The main results and conclusions from this study are
summarized in Sect.~\ref{summary}.

\section{Data and general description \label{data}}
The ETGs discussed here, \object{NGC 1167} (SA0), \object{NGC 1349} (S0) and
\object{NGC 3106} (S0), at a distance of, respectively, 66.2, 87.7 and 90.1
Mpc, according to NED
(\href{https://ned.ipac.caltech.edu/}{https://ned.ipac.caltech.edu/}), were
included in a statistical study by P13 with the main focus on the radial
distribution of the \ewha\ and diagnostic emission-line ratios after \citet[]
[hereafter BPT]{bpt81}, followed by a more detailed 2D analysis of their
nebular and stellar component in G15b. The second brightest member of a group
of six normal-luminosity ($M_B \simeq -20$) galaxies is \object{NGC 1167},
whereas \object{NGC 1349} can be considered truly isolated, since they lack
any bright companion within $\sim$500 kpc, according to NED.  As for
\object{NGC 3106}, it is a field galaxy with seven dwarfish ($-18.5 \la M_r
\la -15.1$) companions listed in NED, with a linear and a velocity separation
of 0.3-10.2 kpc and $\leq$100 \kmsec, respectively.  An in-depth study of
these low-mass companions, two of which display emission-line spectra, will be
presented in Kehrig et al. (in prep.). With the exception of NGC\ 1167,
classified as a Seyfert, as also evident from its prominent radio jet, the
nuclear spectra of these type~i+ ETGs show faint nebular emission
(\ewha$\leq$3\AA) and LINER BPT ratios within their effective radius \reff\
(G15b).  With regard to their photometric structure, the ETGs under study
qualify by their bulge S\'ersic index and bulge-to-total (B/T) ratio ($\eta
\la 2$ and $\leq$1/3, respectively; Breda et al. (in prep.), hereafter B15) as
pseudo-bulge galaxies.  Besides SDSS imaging data, this study uses
low-spectral resolution ($R\sim 850$) IFS CALIFA data with PMAS/PPaK
\citep{Roth05,Kelz06} reduced with the CALIFA data processing pipeline v1.5
\citep{GarciaBenito2015}.  Emission-line flux and equivalent width maps were
computed with the Porto3D pipeline (Kehrig et al. 2012, hereafter K12; P13;
see G15b for a full description of its latest version), which allows for
spaxel-by-spaxel spectral fitting with \SL\ \citep{cid05} and subtraction of
the best-fitting stellar spectrum from the observed one, thereby permitting
accurate extraction of the pure nebular emission.

\section{Spiral-like SF features in the ETG periphery\label{results}}
\begin{figure*}
\begin{picture}(18.4,15.0)
\put(0,10.1){\includegraphics[width=4.75cm, viewport=20 30 520 290]{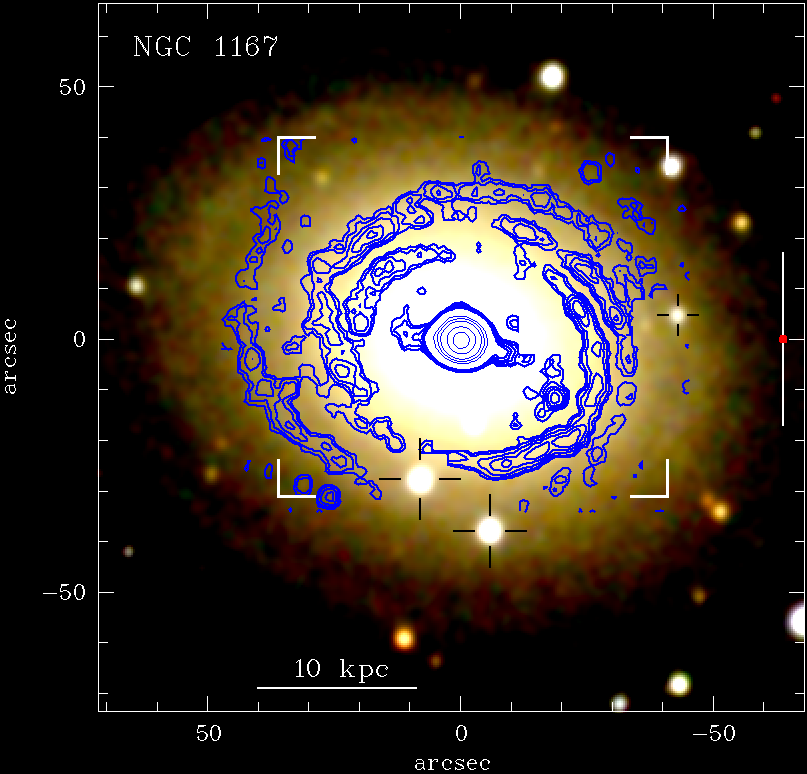}}
\put(6.4,10.1){\includegraphics[width=4.4cm, viewport=20 30 520 290]{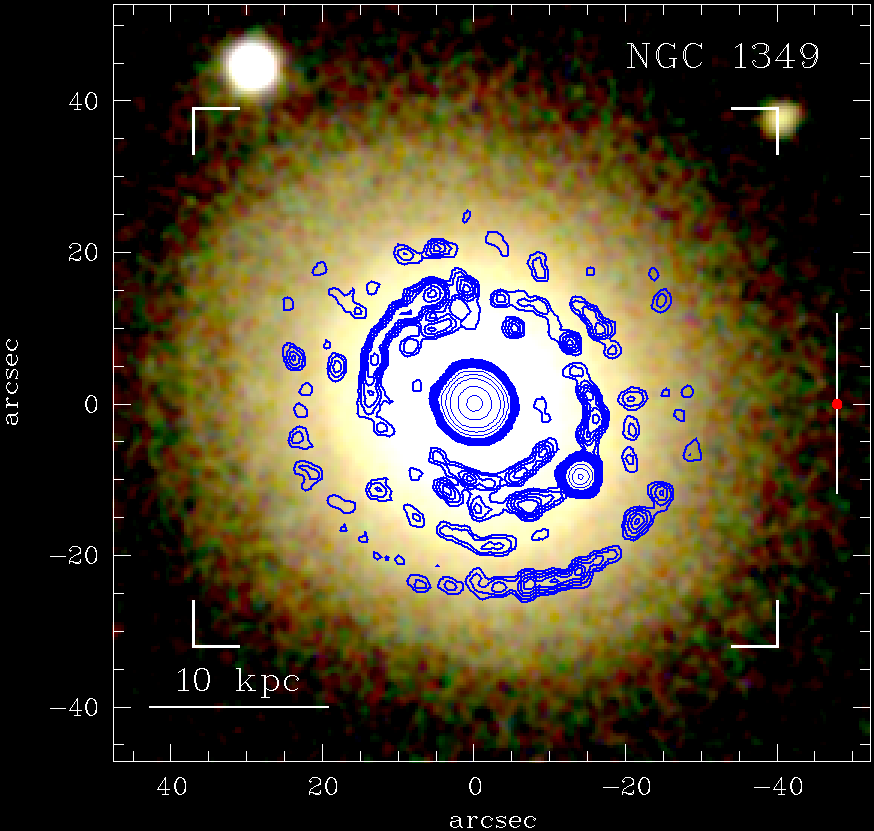}}
\put(12.7,10.1){\includegraphics[width=3.64cm, viewport=20 30 520 290]{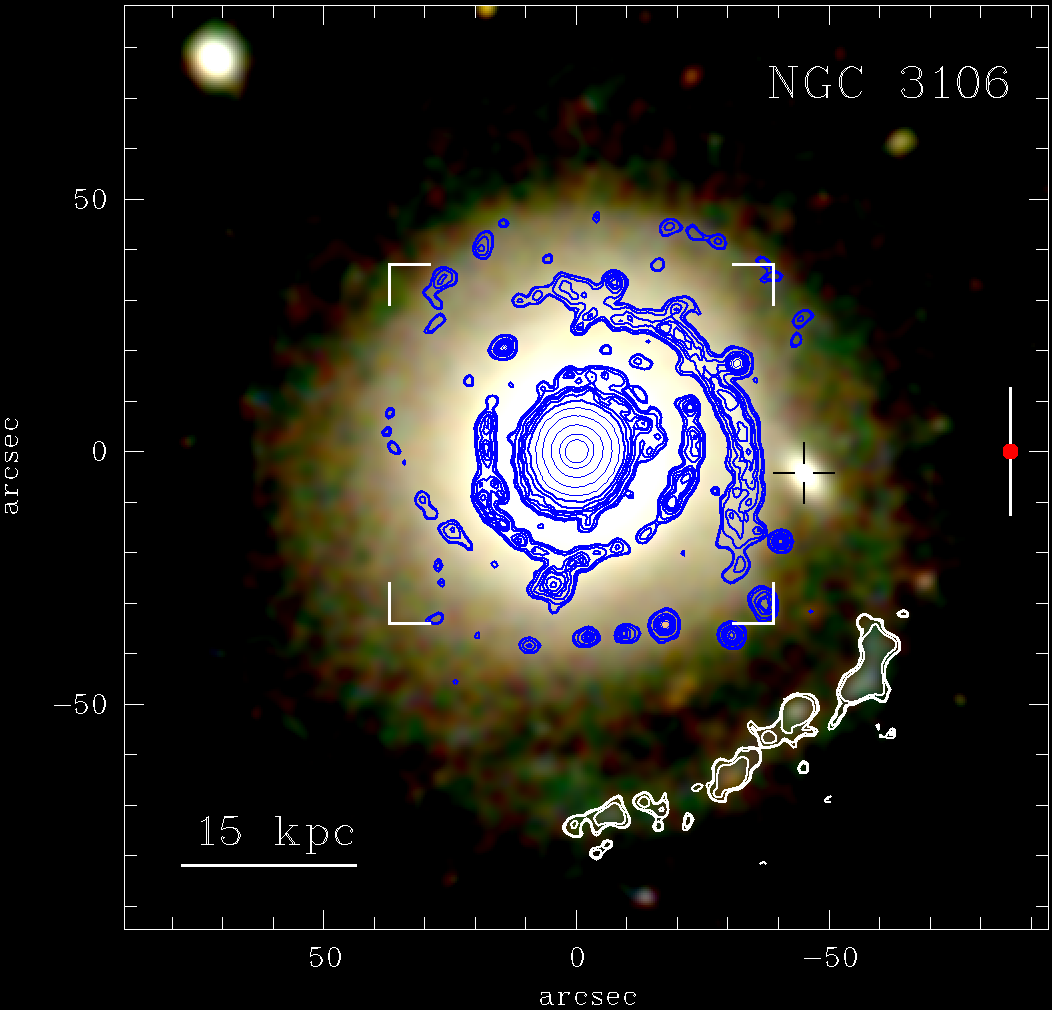}}

\put(0.00,4.7){\includegraphics[width=3.6cm, viewport=20 30 520 290]{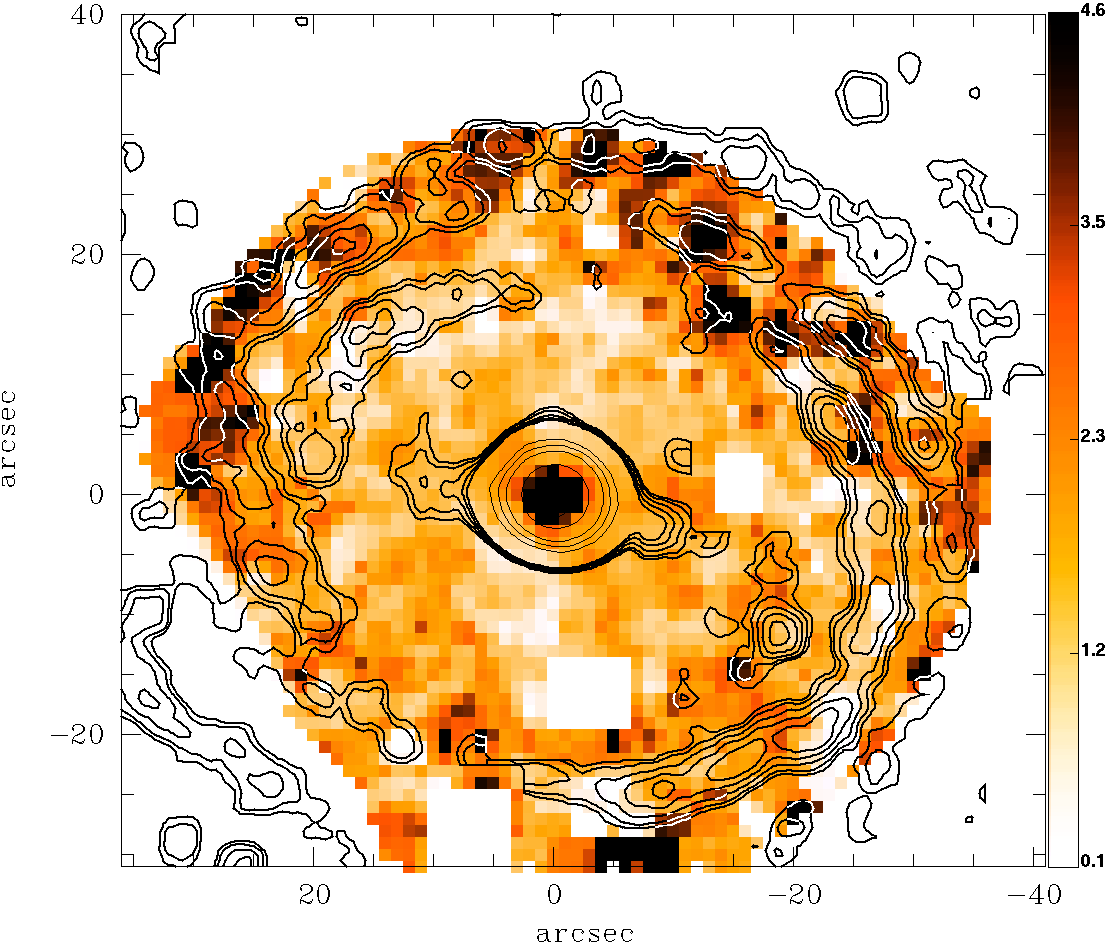}}
\put(6.4,4.7){\includegraphics[width=3.6cm, viewport=20 30 520 290]{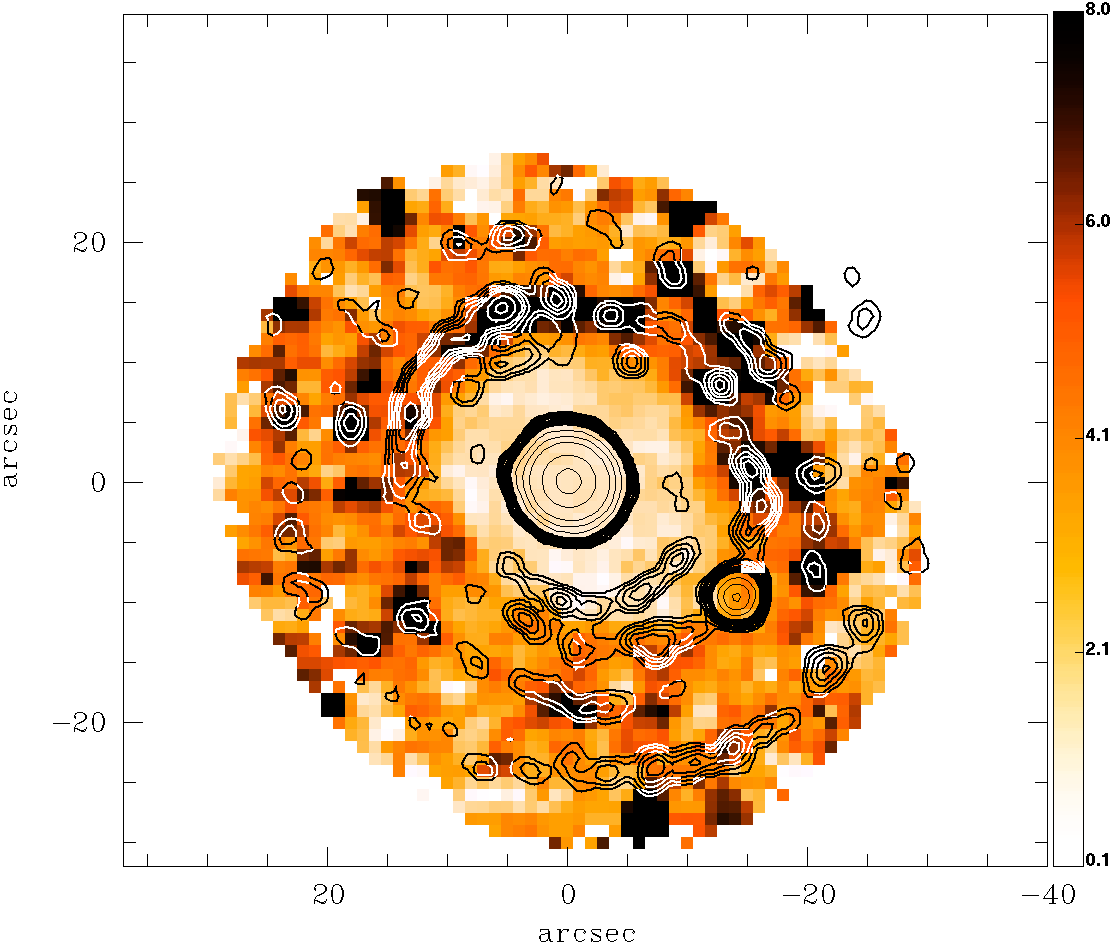}}
\put(12.7,4.7){\includegraphics[width=3.9cm, viewport=20 30 520 290]{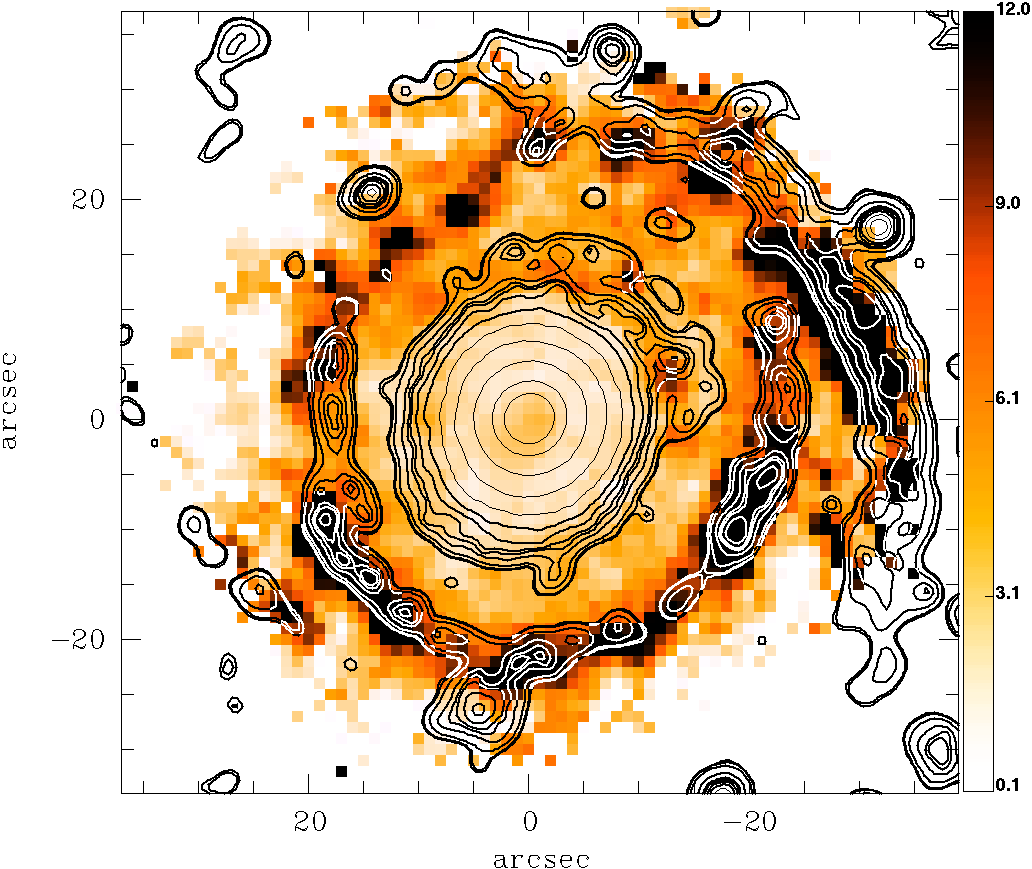}}

\put(00.00,04.30){\includegraphics[width=4.4cm, angle=-90.0]{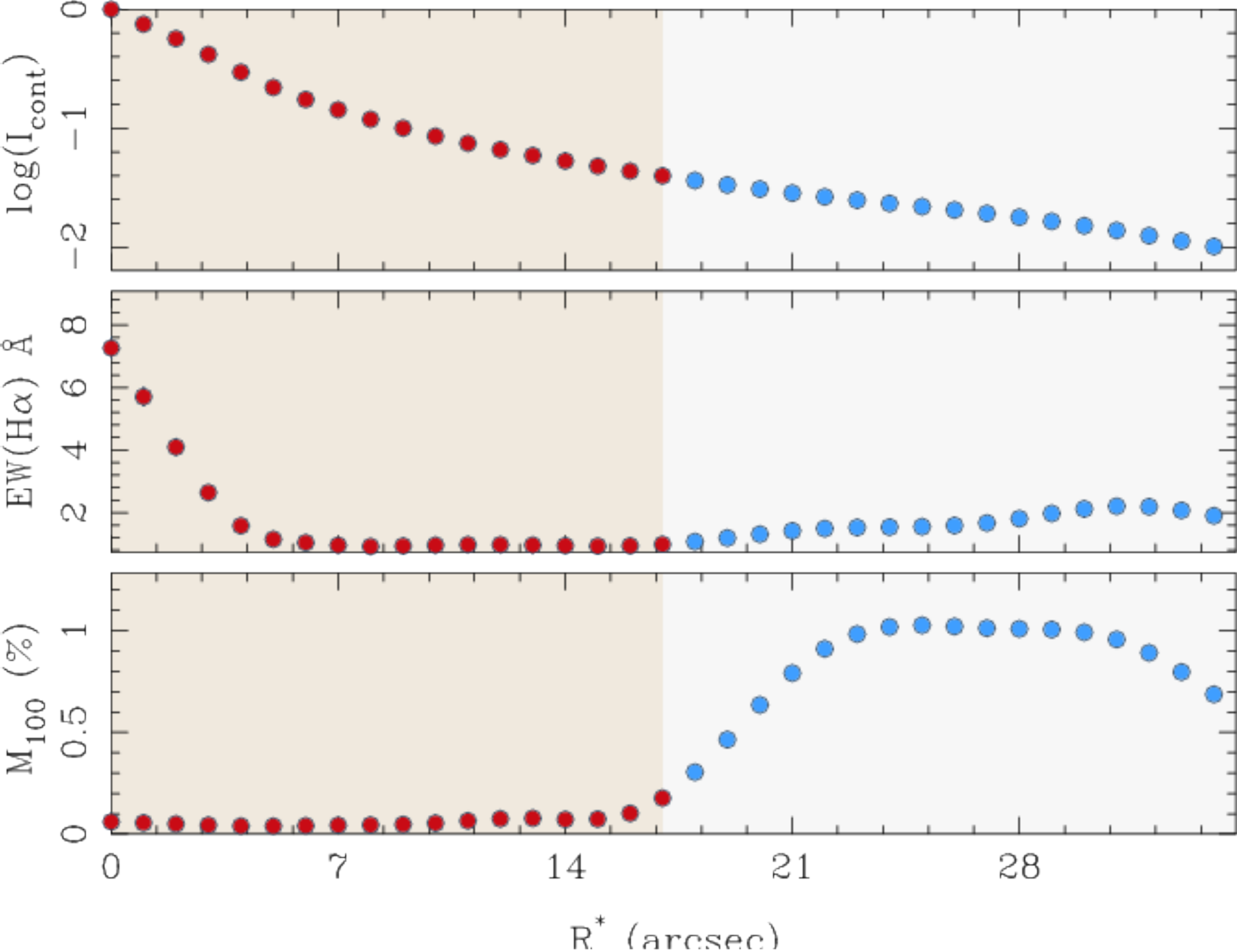}}
\put(06.40,04.30){\includegraphics[width=4.4cm, angle=-90.0]{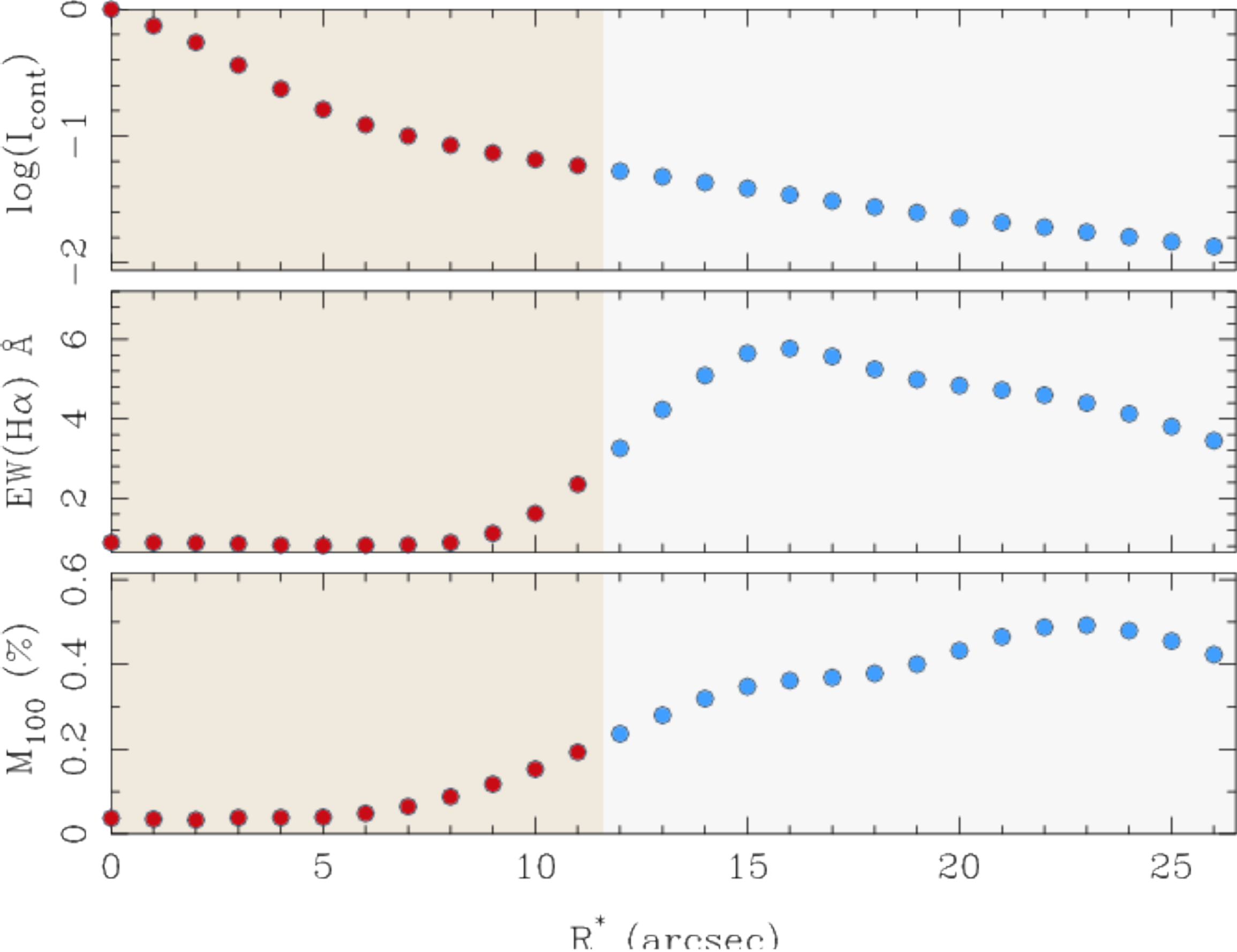}}
\put(12.70,04.30){\includegraphics[width=4.4cm, angle=-90.0]{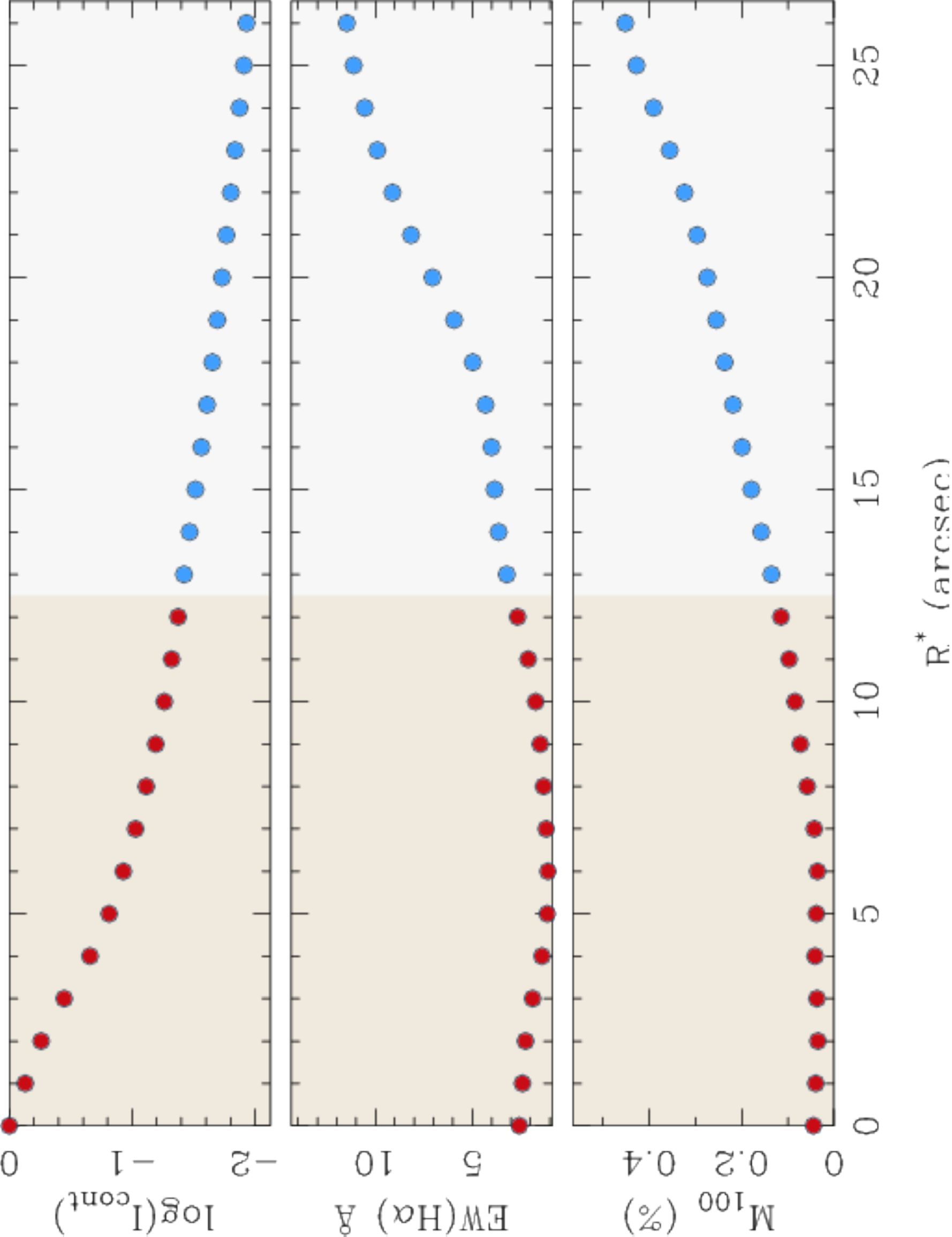}}

\PutLabel{0}{14.8}{\color{white}\mcap a}
\PutLabel{6.4}{14.8}{\color{white}\mcap b}
\PutLabel{12.7}{14.8}{\color{white}\mcap c}

\PutLabel{00.70}{14.75}{\begin{tcolorbox}[boxsep=1pt,colframe=black,bottom=0pt,top=0pt,left=0pt,width=2.0cm,colback=black]\color{white}\vcap NGC 1167\end{tcolorbox}}
\PutLabel{09.94}{14.75}{\begin{tcolorbox}[boxsep=1pt,colframe=black,bottom=0pt,top=0pt,left=0pt,width=2.0cm,colback=black]\color{white}\vcap NGC 1349\end{tcolorbox}}
\PutLabel{16.25}{14.75}{\begin{tcolorbox}[boxsep=1pt,colframe=black,bottom=0pt,top=0pt,left=0pt,width=2.0cm,colback=black]\color{white}\vcap NGC 3106\end{tcolorbox}}

\PutLabel{0}{9.05}{\color{black}\mcap d}
\PutLabel{6.4}{9.05}{\color{black}\mcap e}
\PutLabel{12.7}{9.05}{\color{black}\mcap f}
\PutLabel{0.7}{3.4}{\color{black}\lcap g}
\PutLabel{7.1}{3.4}{\color{black}\lcap h}
\PutLabel{13.4}{3.4}{\color{black}\lcap i}
\PutLabel{5.4}{2.4}{\color{black}\lcap j}
\PutLabel{7.1}{2.4}{\color{black}\lcap l}
\PutLabel{13.4}{2.4}{\color{black}\lcap k}
\PutLabel{0.7}{0.9}{\color{black}\lcap m}
\PutLabel{7.1}{0.9}{\color{black}\lcap n}
\PutLabel{13.4}{0.9}{\color{black}\lcap o}
\end{picture}
\vspace*{-2mm}
\caption[]{{\nlx a--c:} True-color SDSS image composites of the type~i+ ETGs
  under study. The overlaid contours depict the morphology of the spiral-like
  star-forming features, as revealed by applying an unsharp-masking technique.
  The outermost contour corresponds to a surface brightness level $\mu_r\simeq
  25$ \sbb\ for \object{NGC 1167} and $\simeq 26$ \sbb\ for \object{NGC 1349}
  and \object{NGC 3106}, and the vertical line to a linear scale of 2\reff\ in
  the SDSS $r$ band. Bright foreground stars (panel a) and a dwarf galaxy at
  the redshift of NGC\ 3106 (panel c) are shown with crosses, and the edges of
  the CALIFA field of view are delineated in white color.  {\nlx d--f:} \ewha\
  maps, as derived from CALIFA IFS data, overlaid with the same contours as in
  panels a-c.  Radial profiles of the intensity of the stellar continuum,
  normalized to its central value {\nlx(g--i)}, \ewha\ {\nlx (j--k)} and mass
  fraction $M_{100}$ (\%) of stars younger than 100 Myr {\nlx (m--o)}.  }
\vspace*{-2mm}
\label{fig:spiral-features}
\end{figure*}

Figure \ref{fig:spiral-features} gives a synoptic view of the photometric and
spectroscopic evidence for the presence of spiral-like star-forming patterns
in the three i+ ETGs.  Already visual inspection of SDSS poststamp images
reveals traces of faint bluish patches and/or ripples in their low-surface
brightness (LSB) extra-nuclear component.  After a flux-conserving unsharp
masking technique \citep{P98} on SDSS broadband data, these subtle patterns
become clearly discernible as contiguous spiral-like features (contours in
panels a--c) over a linear extent of 20--40 kpc, in close resemblance to
spiral arms in normal late-type galaxies.  Because of their faintness ($24 \la
\mu$ (r \sbb) $\la 26$), the detected structures contribute on average less
than 10\% of the local line-of-sight intensity, thereby leaving only subtle
signatures on color maps and radial color profiles (see B15 for details).
Likewise, their integrated magnitude (--16 $\la M_r \la $ --18) translates to
a negligible ($<$2\%) fraction of the total galaxy luminosity.
%
Whereas in \object{NGC 1349} these features protrude slightly inwards of
\reff\ (11\farcs6) and extend over a radial zone 0.7$\leq$\reff$\leq$2.3, in
the case of NGC~1167 and NGC~3106 they are detected well outside the nuclear
region (1.2--2.6 \reff\ with \reff=17\arcsec\ and 1.5--6.2 \reff\ with
\reff=12\farcs5, respectively).  This is to be contrasted with several
late-type pseudobulge galaxies \citep[e.g.,][]{KormendyKennicutt2004} and the
S0 \object{NGC~6762} where K12 found through application of the same
unsharp-masking technique, faint spiral features to be exclusively embedded
within the pseudobulge ($<$\reff).

In panels d-f of Fig.~\ref{fig:spiral-features} the spiral-like features
revealed by unsharp-masking (contours) are shown overlaid with \ewha\ maps
obtained from CALIFA IFS data.  It can be seen that the former are spatially
correlated with more extended zones of enhanced \ewha, which implies that they
are not purely stellar relics from fading spiral arms that have long ceased
forming stars, but instead sites of ongoing SF.  Evidently, the morphology of
the detected LSB features is hardly compatible with the interpretation of them
arising from shock excitation due to, e.g., infalling gas, or ionization by a
central AGN.  Likewise, the hypothesis of photoinization by the (evenly
distributed) pAGB stellar background can be discarded on morphological
grounds.  Another argument rests on the \ewha\ of these spiral-like
structures, determined to be 2.6--5~\AA\ in \object{NGC 1167} and reaching to
$>$20~\AA\ in \object{NGC 3106}: As shown in, e.g., \citet{bin94},
\citet{cid11} and, in more detail, in G15b, pAGB photoinization can account
for a maximum \ewha$\simeq$2.4~\AA\ ($\equiv$\pAGBmax) for a broad range of
stellar metallicities and IMFs.  The fact that the \ewha\ exceeds \pAGBmax,
rules out pAGB photoionization as the physical origin of the detected
spiral-like features.

The outer rise of the \ewha\ above \pAGBmax\ can also be appreciated from the
radial profiles (panels j-k). These were derived with the isophotal
  annuli (\isan) technique \citep[][see K12, P13 and G15b for applications to
  IFS data]{P02}, whereby the emission-line free stellar continuum between
6390 \AA\ and 6490 \AA\ (panels g-i) was used for the \isan\ definition. It
can be seen that in all cases the mean \ewha\ increases smoothly beyond
\reff\ (shaded area) and reaches in the case of \object{NGC 1349} and
\object{NGC 3106} values between $\sim$6 \AA\ and $\ga$12 \AA.

In summary, the combined analysis of SDSS imaging and CALIFA IFS data
consistently indicates that the LSB spiral-like features in the extra-nuclear
component of the three type~i+ ETGs under study are sites of ongoing SF.  It
should be pointed out that the presence of H{\sc ii} regions in the periphery
of our ETGs has previously been noticed in the course of an investigation of
metallicity gradients in the CALIFA galaxy sample
\citep{Sanchez2013,Sanchez2014}, even though not given a detailed discussion
owing to the scope and statistical nature of those studies.  Also G15b show
that the outer SF zone in type~i+ ETGs appears on BPT diagrams to be very
similar to H{\sc ii} regions in late-type galaxies and it is generally located
in the `intermediate' zone between the empirical SF sequence by
\cite{Kauffmann03} and the `maximum SF' boundary from \cite{Kewley01}.
Obviously, the detectability and spectroscopic BPT classification of such
embedded spiral features depends on their intrinsic properties (\ha\ and
broad-band surface brightness) and the observational data used (e.g.,
spectroscopic aperture, continuum S/N).

Quantitatively, our results in G15b indicate that the SF zone in type~i+ ETGs
is discernible on radial BPT-ratio profiles only when it contains appreciable
nebular emission, as in the case of, e.g., \object{NGC 1349} and \object{NGC
  3106}. The \ha\ luminosity fraction associated with the spiral-like SF
features increases from $\sim$26\% in NGC\ 1167 to 68\% in NGC\ 3106, reaching
up to 80\% in NGC\ 1349.  Despite their faintness on broadband images, they
therefore provide a significant, if not the dominant fraction of the total
\ha\ emission. Their \ha\ luminosity, corrected for the contribution of pAGB
photoionization, translates by the \citet{Kennicutt89} calibration to a SFR
between 0.08 and 0.32 \msun~yr$^{-1}$.  The latter yields for the \mstar\
determined by G15b (1.5\dots 2.6 $\times 10^{11}$ \msun), a very low ($\la
10^{-12}$ yr$^{-1}$) specific SFR. This value will increase by no more than
30\% if intrinsic extinction is taken into account, given the only slightly
enhanced \ha/\hb\ ratio (2.95-3.2) in the star-forming periphery.  Hence, SFRs
based on the \ha\ luminosity, even if sustained over 1~Gyr, would produce only
a tiny ($\ll$1\%) fraction of the present-day \mstar.  It is also noteworthy
that, despite their low SFR surface density ($\la 1 \times 10^{-3}$ \msun\
yr$^{-1}$ kpc$^{-2}$), $\sim$3 times lower than the average value in the Milky
Way and comparable to that in LSB dwarf irregulars
\citep[cf][]{KennicuttEvans2012}, the detected spiral-like features have to
contain a high-mass tail of ionizing stars.  This yields a potentially
important observational constraint in the context of the ongoing debate on
possible stochasticity effects \citep[e.g.,][]{CMM94} and spatial variations
in the IMF of massive ETGs \citep[e.g.,][]{,Cappellari12-IMF}.

A question of considerable interest is whether the LSB spiral-like patterns in
i+ ETGs witness a brief rejuvenation episode that has been triggered by, e.g.,
capture of a gas-rich satellite, or, instead represent localized SF signatures
of a spatially and temporally extended inside-out formation process.  Various
lines of evidence favor the second interpretation.  For example, deep near-UV
archival data with GALEX impressively reveals SF activity all over the
extended LSB periphery of \object{NGC 1167}. The massive ($1.8 \times 10^{10}$
\msun) H{\sc i} halo of this galaxy that reaches out to $\sim$80 kpc
\citep{Emonts2006}, and the presence of extended H{\sc i} reservoirs in about
1/4 of mostly disky ETGs \citep{Serra12} suggest that a significant fraction
of these seemingly `old and dead' systems have the gas supply to sustain SF
and their continued growth.  The ongoing inside-out assembly of type~i+ ETGs
is also indicated by the outwardly increasing mass fraction of stars younger
than 100 Myr ($M_{100}$), which reaches mean values between $\sim$0.4\% and
$\sim$1\% for $r\ga$\reff\ (Fig.~\ref{fig:spiral-features}, panels m-o).
Assuming that in ETGs light roughly scales with stellar mass, these values
translate to an increase by 2\%-5\% in the total \mstar\ over the past 1 Gyr,
in agreement with previous evidence for a significant growth of \mstar\ since
$z\sim 1$ (cf Introduction).  The tension between SFRs estimated from the \ha\
luminosity of the spiral-like features ($< 1$\msun/yr) and those obtained from
spectral synthesis (a few \msun/yr) suggest that Balmer-line emission may
yield a lower bound to the total SFR of ETGs, presumably because of extensive
Lyman continuum photon escape through the tenuous/porous gas component of ETGs
(cf discussion in P13). In addition, a spatially varying IMF, being fully
populated at the locus of the spiral features and truncated at its ionizing
high-mass end elsewhere, might offer a non-mutually exclusive alternative.

It is conceivable that type~i+ ETGs have previously been forming stars at a
much higher pace -- a conjecture in qualitative agreement with the presence of
ionized gas disks in intermediate-redshift ETGs \citep{Jaffe14} -- and are now
seen in an advanced stage of a prolonged, gradually declining SF process
\citep[e.g.,][]{Kaviraj07}.  The Gyr-long inside-out growth of ETGs is also
suggested by the outwardly increasing luminosity contribution of stellar
populations younger than 5~Gyr (G15b) and their by $\sim$0.1-0.25 dex lower
light-weighted age in their periphery \citep[e.g.,][]{Rosa15}, following a
general trend for Hubble-type galaxies \citep[e.g.,][]{Perez13}.  The observed
size evolution of quiescent galaxies at $z \sim 2$ to present-day ETGs
\citep[e.g.,][]{Daddi05,vanDokkum08} is also consistent with a gradual
inside-out assembly process.

Such considerations, and the observational evidence presented here, support
the view that the subtle spiral-like SF features detected in type~i+ ETGs
through unsharp-masking primarily reflect a continuous inside-out
\mstar\ growth process. Even though these systems may experience a transient
rejuvenation through, e.g., accretion of a gas-rich satellite, this process
alone appears incapable of accounting for various observational constraints,
including the smooth radial increase of the mass fraction $M_{100}$ of stars
younger than 100 Myr.

The observed spatial segregation of the nebular component of i+ ETGs in two
spectroscopically distinct radial zones - an inner SF-devoid zone that contains
faint LINER emission and an outer star-forming zone with strong nebular
emission - calls for an examination of possible aperture effects in studies of
such galaxies and their morphological analogs.  For example, \citet{Pracy14}
report the presence of ongoing SF in the outskirts of two E+A post-starburst
galaxies and point out that the previously lacking evidence for
SF-activity in these systems can be explained by an aperture bias in
single-fiber spectroscopic studies. Likewise, an analogous yet inverse
core-envelope spectrophotometric segregation is typical among low-mass
star-forming galaxies (e.g., blue compact dwarfs or `green peas') where
starburst activity is confined to the central part of an older stellar host
\citep[e.g.,][]{P96b,GdP03,Amorin12}.

\section{Summary and conclusions \label{summary}}
We report on the detection of faint ($\mu\simeq$24--26 \sbb), extended (20--40
kpc in diameter) spiral-arm-like features in the periphery (\rr$\ga$\reff) of
three early-type galaxies (ETGs) through application of an unsharp-masking
technique on SDSS imaging data.  An analysis of integral field spectroscopy
data from the CALIFA survey reveals a tight spatial correlation of these
subtle spiral-like features with a broader zone of enhanced
(3\AA$\la$\ewha$\la$20\AA) nebular emission in the ETG outskirts, which
implies that they are not purely stellar relics from fading spiral arms that
have long ceased forming stars, but sites of ongoing low-level star formation
(SF).

The \ha\ luminosity of the detected spiral-like features translates into a star
formation rate of $\sim$0.1--0.3 \msun/yr and a very low
($\sim$$10^{-12}$ yr$^{-1}$) specific SFR.  These estimates are
significantly lower than those obtained from spectral synthesis models, which
indicate that the i+ ETGs under study have assembled 2\%-5\% of their total
stellar mass over the past one Gyr in an inside-out SF process. We argue that
this discrepancy may be due to extensive Lyman continuum photon escape or a spatially 
varying IMF.

Despite their faintness, the detected star-forming spiral-arm-like
features are of considerable interest as they document the still ongoing
growth of seemingly `old and dead' ETGs out of a cold-gas reservoir in
their outskirts and could add valuable observational insights into the origin
and stability of spiral structure in galaxy spheroids.

All but one of the studied ETGs, classified as type~i+, display faint
(\ewha$\simeq$1~\AA) LINER emission in their central part (\rr$\leq$10\arcsec)
and a spatial confinement of up to 80\% of their integral \ha\ emission beyond
the effective radius. This spatial segregation of the nebular component of i+
ETGs in two spectroscopically distinct radial zones -- an inner SF-devoid zone
that contains faint LINER emission and an outer star-forming zone with strong
nebular emission -- calls for an examination of possible aperture effects in
studies of such galaxies, and their morphological analogs, with single-fiber
spectroscopic data. 
\begin{acknowledgements}
 This paper is based on data from the Calar Alto Legacy Integral Field Area
 Survey, CALIFA (http://califa.caha.es), funded by the Spanish Ministery of
 Science under grant ICTS-2009-10, and the Centro Astron\'omico
 Hispano-Alem\'an. JMG acknowledges support by Funda\c{c}\~{a}o para a
 Ci\^{e}ncia e a Tecnologia (FCT) through the Fellowship SFRH/BPD/66958/2009
 and POPH/FSE (EC) by FEDER funding through the program Programa Operacional
 de Factores de Competitividade (COMPETE). PP is supported by FCT through the
 Investigador FCT Contract No. IF/01220/2013 and POPH/FSE (EC) by FEDER
 funding through the program COMPETE. JMG\&PP acknowledge support by FCT under
 project FCOMP-01-0124-FEDER-029170 (Ref. PTDC/FIS-AST/3214/2012), funded by
 FCT-MEC (PIDDAC) and FEDER (COMPETE). SFS acknowledges support from
 CONACyT-180125 and PAPIIT-IA100815 grants. Support for LG is provided by the
 Ministry of Economy, Development, and Tourism's Millennium Science Initiative
 through grant IC120009, awarded to The Millennium Institute of Astrophysics,
 MAS. LG acknowledges support by CONICYT through FONDECYT grant 3140566. CJW
 acknowledges support through the Marie Curie Career Integration Grant
 303912. IM acknowledges financial support by the Junta de Andaluc\'ia through
 project TIC114, and the Spanish Ministry of Economy and Competitiveness
 (MINECO) through projects AYA2010-15169 and AYA2013-42227-P. R.A. Marino is
 funded by the Spanish program of International Campus of Excellence Moncloa
 (CEI). This research made use of the NASA/IPAC Extragalactic Database (NED)
 which is operated by the Jet Propulsion Laboratory, California Institute of
 Technology, under contract with the National Aeronautics and Space
 Administration.
\end{acknowledgements}

\end{document}